# Decelerating Growth of Vertically Aligned Carbon Nanotube Arrays: Kinetic Controlled or Diffusion Controlled?

*Rong Xiang*[a,b]*, Zhou Yang*[a]*, Qiang Zhang*[a]*, Guohua Luo*[a]*, Weizhong Qian*[a]*, Fei Wei*[a]**, Masayuki Kadowaki*[b]*, Shigeo Maruyama*[b*]

[a]Beijing Key Laboratory of Green Chemical Reaction Engineering and Technology, Department of Chemical Engineering, Tsinghua University, Beijing 100084, China

[b]Department of Mechanical Engineering, The University of Tokyo, 7-3-1 Hongo, Bunkyo-ku, Tokyo 113-8656, Japan

[*] To whom correspondence may be addressed.

Fei Wei, weifei@flotu.org, Department of Chemical Engineering, Tsinghua University, Beijing 100084, China; tel, 86-10-62788984; fax, 86-10-62772051.

Shigeo Maruyama, maruyama@photon.t.u-tokyo.ac.jp, Department of Mechanical Engineering, The University of Tokyo, 7-3-1 Hongo, Bunkyo-ku, Tokyo 113-8656, Japan.

ABSTRACT Feedstock and byproduct diffusion in the root growth of aligned CNT arrays was discussed in this work. A non-dimensional modulus was proposed to differentiate catalyst-decay controlled growth deceleration from diffusion controlled one. It was found that aligned MWNT arrays are usually free of feedstock diffusion while SWNT arrays are usually facing strong diffusion limit. The present method can also be utilized to predict the maximum length that CNT forest can grow in certain CVD process.

Vertically aligned CNT arrays grown on flat substrate[1-7], in which all the tubes are of the similar orientation and length, offer an ideal platform to study the CNT growth mechanisms and kinetics. Since 1996[1], various chemical vapor deposition (CVD) methods, including floating catalytic CVD[2], plasma enhanced CVD[3], thermal CVD[4], alcohol catalytic CVD[5], water assisted CVD[6], etc. have been proposed to synthesize aligned multi-walled and single-walled CNT arrays. These processes usually involve different catalysts, carbon sources and operation parameters, result in products with different morphologies and qualities. However, none of the CNT growth in these processes can survive from gradual deceleration and final stop. To understand and thereby to overcome the underlying deactivation mechanisms become one of the most critical steps to grow nano-scale tubes to real macroscopic materials.

Recently, many groups affirmed the bottom growth mode of their vertically aligned CNTs, indicating that the feedstock molecules have to diffuse through the thick CNT forest, reach the substrate where catalysts locate, and then contribute to the CNT growth.[8-11] In this bottom-up growth, a new problem, diffusion limit of the feedstock from the top to the root, arise and become a unique decelerating growth mechanism. This means, there is feedstock diffusion limit, the carbon source concentration at the CNT root should be lower than bulk concentration. Previously, Zhu[12] fitted his experimentally-obtained film thickness with square root of growth time and stated that the growth deceleration is attributed to the strong diffusion difficulties of feedstock to the CNT root. However, Hart[13] claimed later that their growth curve can be well fit to either diffusion limit or catalyst decay, suggesting that only fitting might be not sufficient to clarify diffusion controlled process from a kinetic controlled (catalyst decay) one. Further, if the process is in the transition region, i.e. not completely diffusion controlled, root square fitting is not available anymore. Here, we propose a method using a non-dimensional modulus to evaluate quantitatively the degree of feedstock diffusion limit (no diffusion limit, transit region, and strong diffusion limit). Alcohol catalytic CVD (ACCVD) grown SWNT was used as a typical example of this method and was found to be free of feedstock diffusion. The byproduct back diffusion, which has never been taken into account previously, can also be estimated by the present method. Considering the similar diffusion behavior in different CVD processes, five of the most frequently used systems were also discussed. The results agree well with the currently available experiment phenomenon.

Vertically aligned SWNT were synthesized on Co/Mo dip-coated quartz substrate at 800 °C from ethanol as carbon source, while MWNT arrays were grown on quartz substrate at 800 °C with simultaneous feeding of cyclohexane and ferrocene. Details of the growth processes can be found in our previous work[14,15]. The lengths of as-grown CNT arrays were measured by SEM (JSM-7000 and JSM-7401) and average diameters were measured by TEM (Joel 2010).

Figure 1 presents the bottom-up growth process of aligned CNT arrays. Feedstock molecules (e.g. ethanol in ACCVD) diffuse from the top to the root of dense CNT forest, deposit and form solid CNTs while other gas byproduct (e.g. $H_2O$ or $H_2$) need to diffuse at the opposite direction from the root to the top. Here, we only consider one-dimensional (along the tube axis) diffusion inside CNT forest. The diffusion from the sides of the forest is neglected because of the following two reasons. First, the side diffusion distance, i.e. width of vertically aligned CNT film (one or two inches) is usually much larger than the top diffusion distance, i.e. film thickness (less than several millimeters). Second, side diffusion is probably more difficult due to the higher collision frequency in the anisotropic structure of vertically aligned CNT forest.

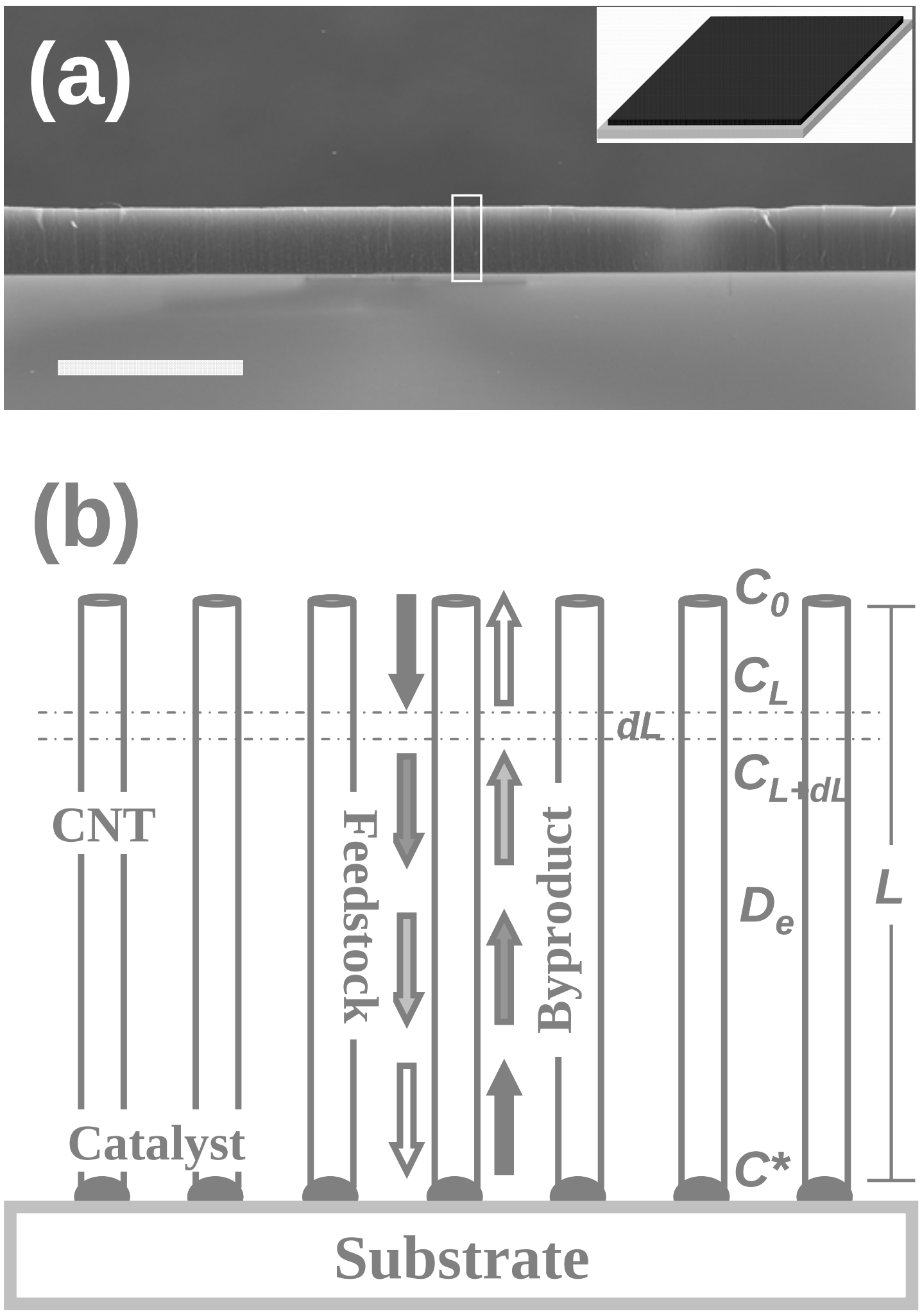

Figure 1. (a) SEM micrograph of vertically aligned SWNT arrays from ACCVD, scale bar 50 μm, inset at top-right is schematic graph of as-growth film on substrate, suggesting the different dimension of film size and thickness; (b) schematic presentation describing the diffusion of feedstock as well as gas product during the bottom-up growth of CNT arrays.

Therefore, in a small sliced CNT forest region *dL* (as indicated by dashes in Fig.1b), the difference in the amount of the feedstock diffusing in from the top and diffusing out from the bottom should be what consumed inside this *dL* region. Then, at CNT-substrate interface, the diffusion flux equals to CNT formation rate (either express by reaction rate $k_sSC^{*m}$ or macroscopic growth rate $aSdL/dt$) when equilibrium. Following the basic diffusion theory, Fick

Law (diffusion flux is proportional to concentration gradient), and reaction theory[16], this process can be expressed by

$$D_e S \frac{dC_{L+dL}}{dL} - D_e S \frac{dC_L}{dL} = 0 \quad \text{(inside of CNT forest)}, \qquad (1)$$

and

$$D_e S \frac{dC_{L+dL}}{dL} - 0 = k_s S C^{*m} = aS \frac{dL}{dt} \quad \text{(root of CNT forest)}, \qquad (2)$$

where $D_e$ is the effective diffusion coefficient, $S$ film area, $L$ length of CNT array, $k_s$ surface reaction constant of carbon source to CNT, $C^*$ effective feedstock concentration at the CNT root, m reaction order, $a$ structure constant of CNT array. Equation (1) suggests that the feedstock concentration is linearly decreasing from top to root, thus, equation (2) can be changed to

$$D_e S \frac{C_0 - C^*}{L} = k_s S C^{*m} \qquad (3)$$

$$D_e S \frac{C_0 - C^*}{L} = aS \frac{dL}{dt}. \qquad (4)$$

Thus as soon as we know reaction order $m$ and reaction coefficient $k_s$, effective concentration $C^*$ can be solved from equation (3) and then time-dependent growth curve can be deduced from equation (4) by integration of $L$ over $t$.

Experiments were carried out under different ethanol pressure in ACCVD to investigate growth order. It is found that the initial growth rate is almost proportional to the concentration (see supporting information), suggesting m=1, which is also found to be approximately valid from the previous report (e.g. for water assisted super growth[17]). If further assuming $k_s$ constant, time dependent growth curve can be deduced from equation (3) and (4) to be

$$L = \sqrt{\left(\frac{D_e}{k_s}\right)^2 + \frac{2D_e C_0}{a} t} - \frac{D_e}{k_s}. \qquad (5)$$

Equation (5) can be proportional to either $t$ (no diffusion limit) or square root of $t$ (strong diffusion limit), depending on the values of $\frac{2D_e C_A}{a} t$ and $\frac{D_e}{k_s}$ (see supporting information). It is similar to what is widely used in silicon oxidation, so-called "Grove-Deal" relationship[18], as discussed before by Zhu[12] and Zhong[19]. One can, in principle, also predict growth curve of CNT array provided that we can get all the parameters listed above. However, a big difference between growth of CNT array and silicon oxide is that, in most of the cases, the catalyst for CNT growth can't survive from poisoning. Therefore, $k_s$ in CNT growth is also a time-dependent parameter, unlike in silicon oxidation where $k_s$ is always constant.

To enable a simple estimate on the existence of diffusion limit for a certain system and CNT length, we can define a no-dimensional factor $\varphi$ by

$$\varphi = \frac{\mathrm{L}}{\mathrm{A}} = \frac{k_s L}{D_e} \qquad (6)$$

It stands for the ratio of diffusive capability to reactive capability. Then, the ratio of effective concentration and bulk concentration $\eta$ (usually called effective factor) can be correlated with $\varphi$ via a simple function from equation (3) as

$$\eta = \frac{C_A^*}{C_A} = \frac{D_e}{k_s L + D_e} = \frac{1}{\varphi + 1}. \qquad (7)$$

This allows us quantitatively characterize the degree of diffusion limit. When $\varphi$ is small (<0.1), i.e. it is much easier to diffuse than to react, the effective index will be nearly 1 (>0.9), indicating that there is small diffusion difficulties. In the contrast condition, when $\varphi$ is very large (>9), i.e.

it is more difficult to diffuse than to react, the effective factor will be nearly zero (<0.1) and the overall reaction will be dominated by diffusion rate. The in-between situation is what we mentioned before as transition region, where the growth curve will be proportional to neither $t$ nor $t^{1/2}$.

In ACCVD, the top of CNT arrays is frequently refreshed and therefore the byproduct concentration can be treated as zero due to the high ethanol flow rate. In this case, the byproduct concentration at the CNT root can also be revealed as a single function of $\varphi$ as

$$C_B^* = C_{A0} \times \sqrt{\frac{M_B}{M_A}} \times \left( \frac{\varphi}{\varphi + 1} \right), \tag{8}$$

if we assume that one $C_2H_5OH$ molecule produces one byproduct molecule, e.g. $H_2O$, after decomposition ($A \rightarrow CNT + B$). The details will be discussed later.

According to the above discussion, as long as we know $D_e$ and $k_s$, the influence of diffusion can be concluded simply from the value of $\varphi$ for a certain CNT length $L$. We know that the average diameter of SWNT in ACCVD is about 2 nm, and the real density of the as-grown film is about 0.041 g/cm$^3$. Therefore the average distance of adjacent CNTs can be easily calculated to be 8.8 nm. As the mean free path of ethanol in this process is about 16000 nm, much larger than the distance between SWNTs, it can be concluded that the ethanol diffusion difficulties are mainly due to the ethanol-CNT collisions, i.e. in the range of Knudsen diffusion. Thereby, the diffusion coefficient can be estimated from current collision theory if assuming CNT tortuosity as diffusion channel tortuosity[20]. As to $k_s$, we can use the initial value at $t=0$ when the CNT growth is free of diffusion limit. With the estimated $D_e$ and experiment-derived $k_s$, $\varphi$ is calculated to be 0.054 (<<1) for 30 μm SWNT arrays in ACCVD. This means the ethanol concentration at the CNT root where the catalyst locate is almost the same (95% from equation 7) as the concentration at the CNT top. Thus, this process is kinetic controlled rather than diffusion controlled. After we peel as-grown film off the substrate, most of the catalysts remain on the substrate, but the substrate is not active for a second growth. This confirmed that the catalyst poisoning contributed to the growth deceleration, which is consistent with above calculation of $\varphi$. We know $H_2O$ is the byproduct of ethanol decomposition, estimating through equation (8) reveals that there are hundreds ppm of water at the CNT root. Considering the previous report on the critical role of $H_2O$ or $O_2$ on the growth of SWNT[6, 21], we plot the concentration distribution of $H_2O$ in Fig. 2f. This result is interesting but currently we are not sure if this water concentration is critical for the SWNT successful nucleation, or it may cause the rapid catalyst deactivation in ACCVD at the same time. Further work is needed.

One may notice the above discussion on the feedstock diffusion is versatile and valid for all the 1$^{st}$ order growth of aligned CNTs, no matter for SWNT or MWNT. Therefore, with the available data in the literature, we are able to estimate the diffusion degree in other CVD processes growth of aligned CNT forest. The only difference here is when estimating the effective diffusion coefficient for MWNT arrays the molecular diffusion should also be taken into account because the mean free path is comparable to the inter-tube distance for MWNT arrays. We choose four other CVD processes 2 mm MWNT array from floating CVD[15,22] (F-MWNT) 2.5mm SWNT array from super growth by Hata et al.[6, 17, 23] (S-SWNT), 2.5 mm SWNT array from microwave plasma CVD by Zhong et al.[7, 19, 24] (P-SWNT) and 400 μm MWNT from Thermal CVD by Zhu et al.[11, 12, 25] (T-MWNT) and compare the result with our 30 μm SWNT from ACCVD (A-SWNT) in Tab. 1. and Fig. 2. It is suggested that mm-scale SWNTs are suffering from strong feedstock diffusion ($\eta < 10\%$) but diffusion limit seems not to be the dominate reason for the decreasing growth of MWNT array, as the concentration at the root of array is of little difference from the bulk concentration, above 90% even when there is no catalyst deactivation (if considering decrease of $k_s$ in real system, the concentration would be higher). However, for mm scale SWNT arrays, φ is usually much larger than 1 and, even there is no catalyst poisoning, the growth rate of mm scale SWNT arrays will still drop to only 10% due to the strong feedstock diffusion difficulty. Thus φ can suggest the maximum length certain process can reach.

Table 1. System parameters and as-calculated $\varphi$ and $\eta$.

| | | A-SWNT | S-SWNT | P-SWNT | T-MWNT | F-MWNT |
|---|---|---|---|---|---|---|
| T | (K) | 1073 | 1023 | 873 | 1023 | 1073 |
| M | (-) | 46 | 28 | 16 | 28 | 84 |
| Density | (g/cm$^3$) | 0.041 | 0.037 | 0.067 | 0.014 | 0.082 |
| Dia | (nm) | 2 | 3 | 2 | 10 | 29 |
| N | (m$^{-2}$) | 8.5E15 | 5.2E15 | 1.4E16 | 3E14 | 2.1E13 |
| $\rho$ | (-) | 0.973 | 0.963 | 0.956 | 0.976 | 0.986 |
| Dadj | (nm) | 8.8 | 10.9 | 6.5 | 48 | 189 |
| Mfp | (nm) | 16000 | 206 | 5500 | 196 | 91 |
| Rate | (m/s) | 2E-7 | 3.75E-6 | 5E-8 | 1.2E-6 | 5E-7 |
| $k_s$ | (m/s) | 2.4E-3 | 9.2E-3 | 5.7E-3 | 1.2E-4 | 3.5E-4 |
| $D_e$ | (cm$^2$/s) | 0.013 | 0.020 | 0.015 | 0.085 | 0.169 |
| $L$ | (mm) | 0.03 | 2.5 | 2.5 | 0.4 | 2 |
| $\varphi$ | (-) | 0.054 | 11.3 | 9.7 | 0.0057 | 0.042 |
| $\eta$ | (-) | 0.949 | 0.081 | 0.093 | 0.994 | 0.960 |

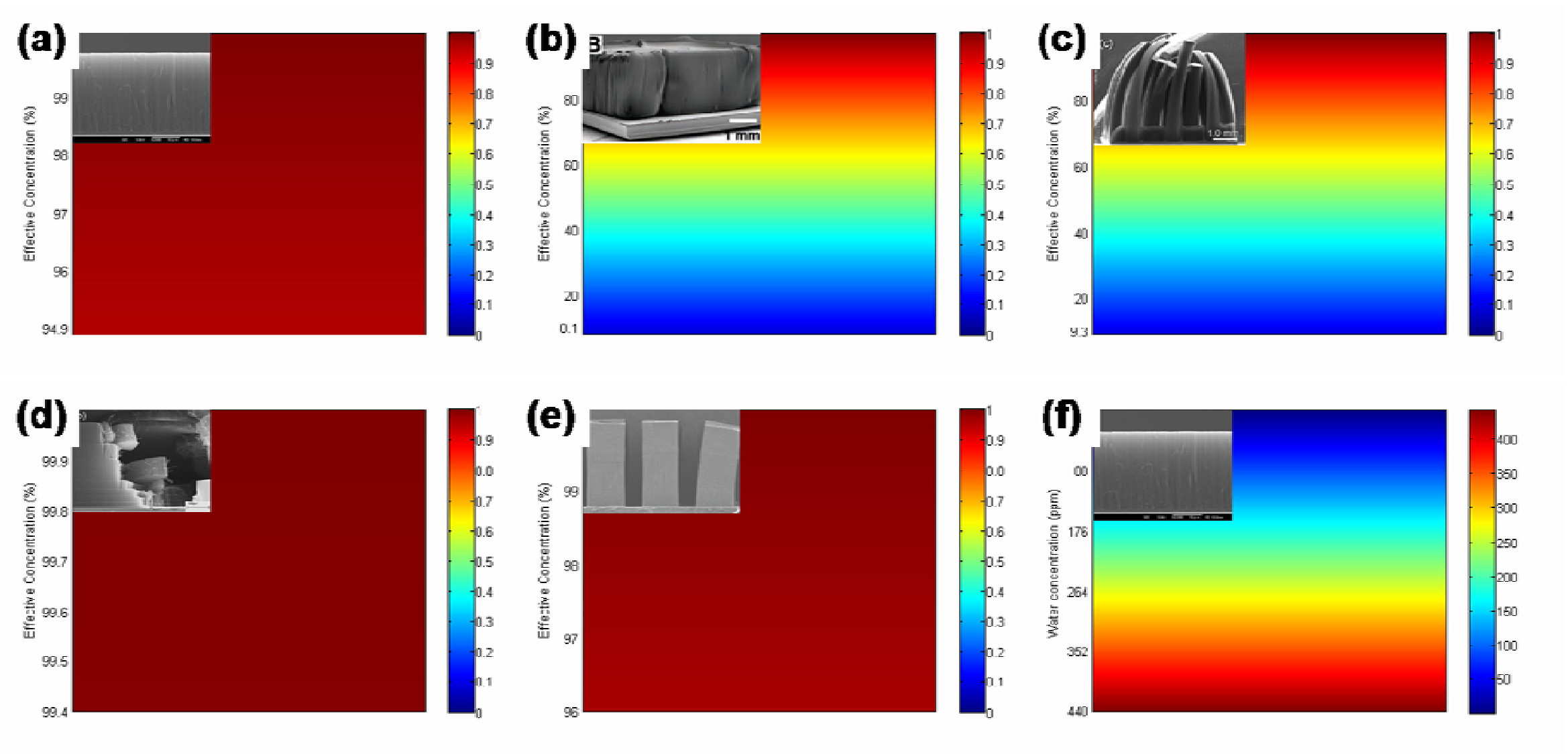

Figure 2. Feedstock or by product concentration distribution in various vertically aligned CNT arrays: (a) 30 um SWNT from ACCVD, (b) 2 mm MWNT from floating CVD, (c) 2.5mm SWNT from water-asisited super growth; (d) 2.5 mm SWNT from microwave plasma CVD, (e) 400um MWNT from thermal CVD; (f) water concentration inside 30 um SWNT from ACCVD. The color in these figures represents the relative/real concentrations.

From equation (6), the influences of different parameters on the diffusion behavior can be investigated and strategies to overcome the diffusion limit for the SWNT growth can be revealed. Larger $D_e$, small $k_s$ or $L$ are all possible ways to decrease $\varphi$. However, the influences of these parameters are very limited because to bring diffusion limited process to reaction controlled region usually needs to decrease $\varphi$ by two orders of magnitude, as expressed in Equation (7). One promising approach is to shorten the diffusion distance by reducing the film size, e.g. making pillar-like or sheet-like patterns of micron scale to allow easy side diffusion. Zhong et al.[13] did this and succeeded in preparing longer 5mm SWNT by making line patterns, in which feedstock can reach the catalyst with much less difficulty than the dense and continuous film. Our calculation result of their system (Fig. 2c) also suggest strong diffusion limit for their process. One may also notice the edge of CNT array grown from “super growth” is usually higher than the center part, which is also an evidence for the diffusion limit in this process.

As to the error in this calculation, it is unavoidable since the influences of some factors, e.g. the bundle structure of SWNTs, the conversion rate of feedstock to CNT, tortuosity of diffusion channel (we assume it to be 1.5 in all cases) are simplified or excluded in the above discussions. However, as mentioned above, error within one order of magnitude in estimate $\varphi$ will not lead to big difference in concluding the existence of diffusion limit. As the biggest error lies on the calculation of $D_e$, further work on direct measurement of $D_e$ is undergoing. Nevertheless, $\varphi$ is helpful in understanding the role of growth parameters on the diffusion limit and the different diffusion behaviors insider SWNT and MWNT arrays

To conclude, here we present the versatile model for the one-dimensional diffusion during the bottom-up growth of aligned CNT arrays. The proposed non-dimensional modulus can be used to evaluate the degree of the diffusion limit of feedstock, as well as byproduct molecules, quantitatively. The results show that, for mm-scale SNWT arrays, the feedstock concentration at root of array is much lower than bulk concentration while for the mm-scale MWNT the diffusion limit can not be attributed to the decreasing growth. The results generated from the model and the possible strategy thereby indicated agrees well with the experiment data.

**Acknowledgement.** The work was supported by China National ‘863’ Program (No. 2003AA302630), China National Program(No. 2006CB0N0702), NSFC Key Program (No. 20236020), FANEDD (No. 200548), Key Project of Chinese Ministry of Education (No. 106011), THSJZ, and National center for nanosicence and technology of China (Nanoctr).

**Supporting Information Available:** Growth curves of aligned SWNT arrays from alcohol CVD and aligned MWNT arrays from floating CVD (Figure S1), evidence for the $1^{st}$ order reaction from ethanol to CNT in alcohol CVD (Figure S2), further explanation on equation (5), details of the calculation, and some more discussion. This material is available free of charge via the Internet at http://pubs.acs.org.

*Supporting information*

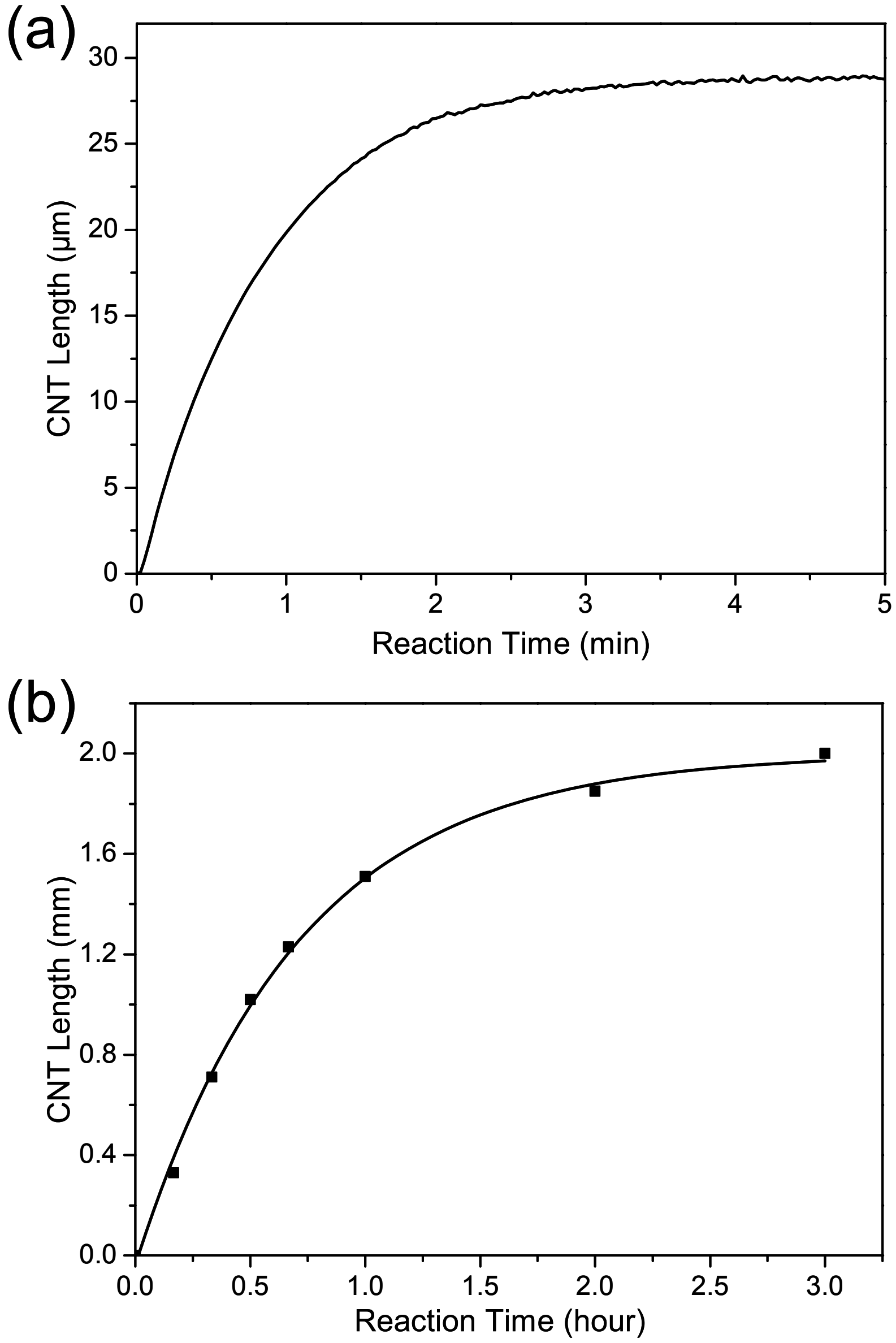

Figure S1. Time dependent growth of vertically aligned (a) SWNT arrays from ACCVD and (b) MWNT arrays from floating CVD, both of which show decelerating growth behaviors over time.

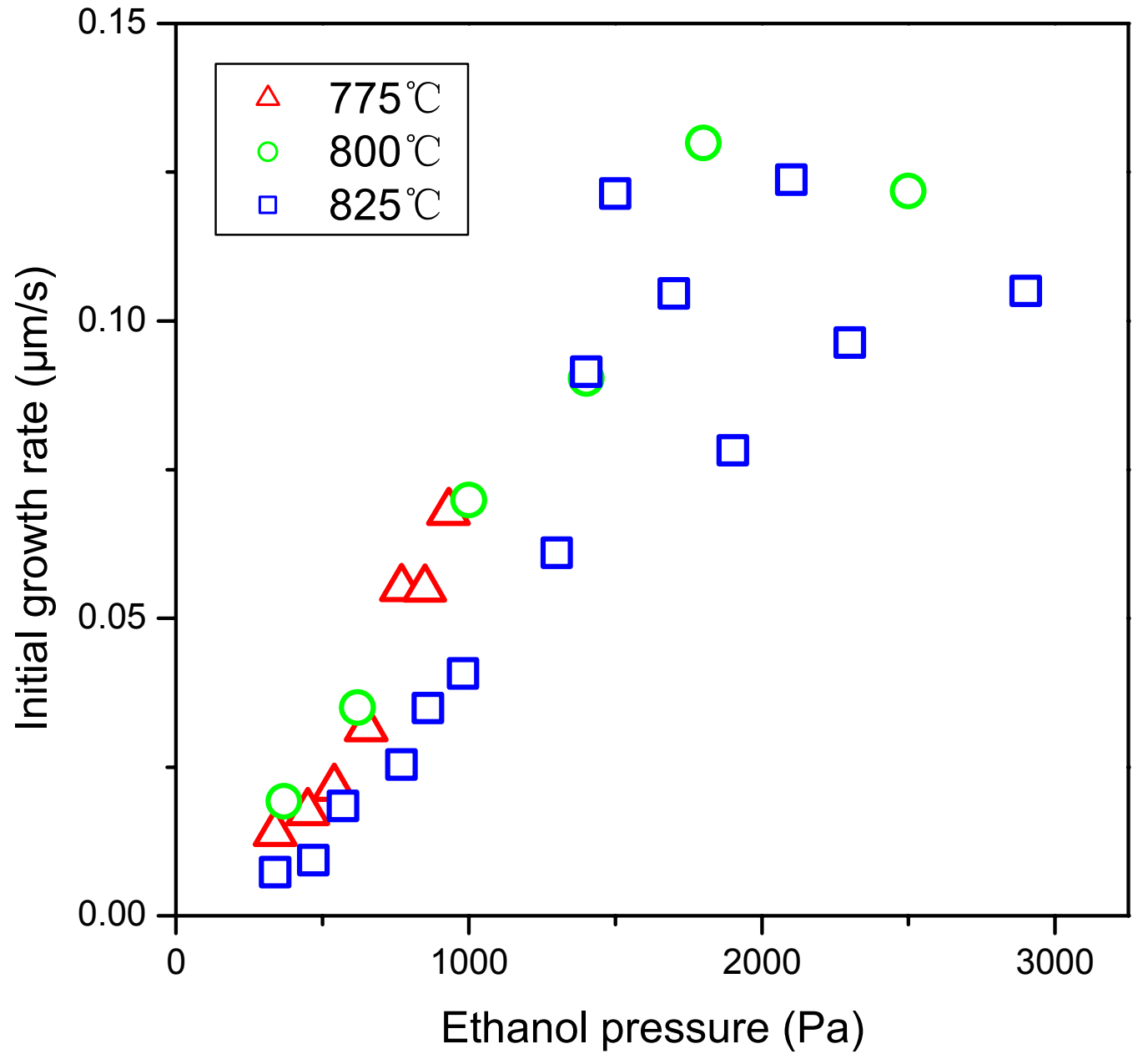

Figure S2. Relationship of initial growth rate of aligned SWNT arrays in ACCVD and the feedstock (ethanol) pressure, confirming the approximate 1st order growth under different temperatures.

Further explanation on equation (5)

Depending on the values of $\left(\frac{\rho D_e}{k_s}\right)^2$ and $\frac{2\rho D_e C_{As}}{a}t$,

$L = \sqrt{\left(\frac{\rho D_e}{k_s}\right)^2 + \frac{2\rho D_e C_{As}}{a}t} - \frac{\rho D_e}{k_s}$ can be either proportional to $t^{1/2}$.

When $\left(\frac{\rho D_e}{k_s}\right)^2 << \frac{2\rho D_e C_{As}}{a}t$,

$$L = \sqrt{\left(\frac{\rho D_e}{k_s}\right)^2 + \frac{2\rho D_e C_{As}}{a}t} - \frac{\rho D_e}{k_s} = \sqrt{\frac{2\rho D_e C_{As}}{a}t} \propto t^{1/2};$$

When $\left(\frac{\rho D_e}{k_s}\right)^2 >> \frac{2\rho D_e C_{As}}{a}t$

$$L = \left(\sqrt{\left(\frac{\rho D_e}{k_s}\right)^2 + \frac{2\rho D_e C_{As}}{a}t} - \frac{\rho D_e}{k_s}\right) \times \frac{\sqrt{\left(\frac{\rho D_e}{k_s}\right)^2 + \frac{2\rho D_e C_{As}}{a}t} + \frac{\rho D_e}{k_s}}{\sqrt{\left(\frac{\rho D_e}{k_s}\right)^2 + \frac{2\rho D_e C_{As}}{a}t} + \frac{\rho D_e}{k_s}} = \frac{C_{As}k_s}{a}t \propto t$$

## Details of some calculation

Mean free path of molecules:

$$\lambda = \frac{RT}{\sqrt{2}\pi d^2 N_A p}$$

where $R$ real gas constant, $T$ temperature, $d$ molecule diameter, $N_A$ Avogadro Constant, $p$ pressure

Knudsen diffusion coefficient:

$$D_K = 9700r\frac{\rho}{\tau}\left(\frac{T}{M}\right)^{1/2}$$

where $r$ is channel diameter, $\rho$ is porosity of CNT membrane, $\tau$ tortuosity of diffusion channel, $T$ temperature, $M$ molecular weight. Tortuosity $\tau$ in our estimation was approximated to 1.5 in all cases because it is the typical value for MWNTs in aligned array. As discussed in the main text it will not bring to much error in main conclusions, i.e. judging the existence (or not) of diffusion limit from φ.

Molecular diffusion coefficient:

$$D_{AB} = 0.001858T^{3/2}\frac{(1/M_A + 1/M_B)^{1/2}}{P\sigma_{AB}{}^2\Omega_{AB}}\left(\frac{T}{M}\right)^{1/2}$$

where T temperature, M molecular weight, P pressure, σ mean molecular diameter, Ω collision integration. A and B stand for two components in gas mixture, which are carbon source (e.g. $C_2H_4$ or $C_6H_{12}$) and carrier gas (e.g. Ar) in our calculation.

Effective diffusion coefficient:

$$D_e = \left( \frac{1}{D_K} + \frac{1}{D_{AB}} \right)^{-1}$$

where $D_k$ Knudsen diffusion coefficient and $D_{AB}$ molecular diffusion coefficient.

## Further discussion on $k_s$

In equation (5)

$$L=\sqrt{\left(\frac{D_e}{k_s}\right)^2+\frac{2D_eC_0}{a}t}-\frac{D_e}{k_s},$$

constant $k_s$ (no catalyst deactivation) is required to fit/predict the time-dependent growth curve. However, in most of the cases, catalyst activity is always diminishing, which means, this equation is too realistic to be applied into real system.

In the present method, initial reaction constant at t=0 (when there is no diffusion problem involved) $k_{s0}$ was sufficient to exclude the diffusion limit for MWNT arrays and predict the CNT length for SWNT arrays. No complete information of catalyst decay is needed.

For the growth of mm-scale MWNT array as shown in figure 2ef, even when constant $k_{s0}$ was used, φ is small and $\eta$ is near 1. If there is some catalyst deactivation along time ($k_s$ will be smaller), φ will be even smaller. Therefore, even when the diffusion is maximized, there is no limit for these aligned MWNT arrays.

For the growth of mm-scale SWNT arrays as shown in figure 2 cd, when $k_{s0}$ is used, the result means, even there is no catalyst deactivation, the growth will be slowed down by the feedstock diffusion. Therefore, in real case with catalyst decay, CNT arrays can never grow over several mm.